\documentclass{aa}
\usepackage{graphicx}
\usepackage{rotating}
\usepackage{txfonts}

\def\bv0{$(B-V)_{0}$}

\def\Msun{$M_{\odot}$ }
\def\Lsun{$L_{\odot}$}
\def\Rsun{$R_{\odot}$ }
\def\Teff{$T_{\rm eff}$ }

\def\vsini{$v\cdot\sin{i}$ }
\def\cind{{\sc Cinderella}}
\def\SigSpec{{\sc SigSpec}}
\def\hd{\mbox{HD\,142666}}

\begin{document}

\title{{\it MOST}\thanks{Based on data from the {\it MOST} satellite, a Canadian Space Agency mission, jointly operated by Dynacon Inc., the University of Toronto Institute for Aerospace Studies and the University of British Columbia with the assistance of the University of Vienna.} photometry of the enigmatic PMS pulsator HD\,142666}

\author{K. Zwintz\inst{1} \and
T. Kallinger\inst{1} \and
D. B. Guenther\inst{2} \and
M. Gruberbauer\inst{1} \and
D. Huber\inst{1,8} \and
J. Rowe\inst{3,4} \and
R. Kuschnig\inst{1,3} \and
W. W. Weiss\inst{1} \and
J. M. Matthews\inst{3} \and
A. F. J. Moffat\inst{5} \and
S. M. Rucinski\inst{6} \and
D. Sasselov\inst{7} \and
G. A. H. Walker\inst{3} \and
M. P. Casey\inst{2}
}

\offprints{K. Zwintz, \\ \email{zwintz@astro.univie.ac.at}}

\institute{Institut f\"ur Astronomie, Universit\"at Wien,
    T\"urkenschanzstrasse 17, A-1180 Vienna, Austria \\
    \email lastname@astro.univie.ac.at  \and
Department of Astronomy and Physics, St. Mary's University, Halifax,
    NS B3H 3C3, Canada \\
    \email guenther@ap.smu.ca, mcasey@ap.smu.ca \and
Department of Physics and Astronomy, University of British Columbia,
    6224 Agricultural Road, Vancouver, BC V6T 1Z1, Canada \\
    \email rowe@phas.ubc.ca, kuschnig@astro.ubc.ca, matthews@astro.ubc.ca, gordonwa@uvic.ca \and
NASA-Ames Research Park, MS-244-30, Moffett Field, CA 94035 \and
D\'epartment de physique, Universit\'e de Montr\'eal, C.P. 6128,
    Succ. Centre-Ville, Montr\'eal, QC H3C 3J7, Canada \\
    \email moffat@astro.umontreal.ca \and
David Dunlap Observatory, University of Toronto, P.O. Box 360,
    Richmond Hill, ON L4C 4Y6, Canada \\
    \email rucinski@astro.utoronto.ca \and
Harvard-Smithsonian Center for Astrophysics, 60 Garden Street,
    Cambridge, MA 02138, USA \\
    \email sasselov@cfa.harvard.edu \and
Institute of Astronomy, School of Physics, University of Sydney, NSW 2006, Australia
    \email dhuber@physics.usyd.edu.au}

\date{Received / Accepted }

\abstract
{Modeling of pre-main sequence (PMS) stars through asteroseismology of PMS p-mode pulsators has only recently become possible, and spacebased photometry is one of the important sources of data for these efforts. We present precise photometry of the pulsating Herbig Ae star \hd\,\,obtained in two consecutive years with the MOST (Microvariability \& Oscilations of STars) satellite.}
{Previously, only a single pulsation period was known for \hd. The MOST photometry reveals that \hd\,\,is multi-periodic. However, the unique identification of pulsation frequencies is complicated by the presence of irregular variability caused by the star's circumstellar dust disk. The two light curves obtained with MOST in 2006 and 2007 provided data of unprecedented quality to study the pulsations in \hd\,\,and also to
monitor the circumstellar variability.}
{Frequency analysis was performed using the routine \SigSpec\,\,and the results from the 2006 and 2007 campaigns were then compared to each other with the software \cind\,\,to identify frequencies common to both light curves. The correlated frequencies were then submitted to an asteroseismic analysis.}
{We attribute 12 frequencies to pulsation. Model fits to the three frequencies with the highest amplitudes lie well outside the uncertainty box for the star's position in the HR diagram based on published values. Some of the frequencies appear to be rotationally split modes.}
{The models suggest that either (1) the published estimate of the luminosity of \hd, based on a relation between circumstellar disk radius and stellar luminosity, is too high and/or (2) additional physics such as mass accretion may be needed in our models to accurately fit both the observed frequencies and \hd's position in the HR diagram.}

\keywords{Stars: pre-main sequence, (Stars:variables:) $\delta$ Sct, Stars:individual:HD 142666, Techniques: photometric}

\maketitle
\titlerunning{MOST photometry of \hd}
\authorrunning{K. Zwintz et al.}

\section{Introduction}
A star evolving from the birthline to the zero-age main sequence (ZAMS) derives most of its energy (half of which heats the star and half of which radiates away) from the release of gravitational potential energy as the star collapses.  These pre-main sequence (PMS) objects are characterized by observational features typical for their early evolutionary stage, such as emission lines, infrared and/or ultraviolet excesses, (ir)regular brightness variations, etc.  The resulting variability of PMS stars is observed on a wide range of time scales and amplitudes.  Variations on time scales of weeks with amplitudes of a magnitude and more originate from the circumstellar environment that contains gas and dust remnants of the stellar birth cloud. Accretion or chromospheric activity induces variations on time scales from several hours to days. Oscillations with periods of few hours to half an hour and amplitudes at the millimagnitude level are due to $\delta$-Scuti-like pulsations in those PMS stars with the right combination of mass, temperature and luminosity.

There are two distinct classes of PMS objects: T Tauri stars and Herbig Ae/Be stars. The low-mass ($<$ 1$M_{\odot}$) T Tauri stars are too cool (spectral types from late F to M) to be unstable to $\delta$-Scuti-type p-mode pulsation. The Herbig Be stars are too massive (4 to 10 $M_{\odot}$) for an observable PMS phase: it is believed that by the time they become first visible optically, they are already burning hydrogen in their cores. The intermediate-mass (1.5 to 4 M$_{\odot}$) Herbig Ae stars - on the other hand - cross the instability region in the Hertzsprung Russell (HR) diagram on their way to the main sequence. They have the right combination of mass, temperature and luminosity to be pulsationally unstable.

Several pulsating PMS stars have been discovered and analyzed within the last few years.  A summary of their properties is given by Zwintz (\cite{zwi08}). The periods of PMS pulsators are in the domain of classical $\delta$ Scuti stars, i.e., between 20 minutes and 6 hours, and their amplitudes lie at the millimagnitude level.  Pre- and post-main sequence stars of the same mass, effective temperature and luminosity mostly differ in their inner structures, but their atmosphere properties are quite similar (Marconi \& Palla \cite{mar98}). PMS stars lack regions of already processed nuclear material and seem not to be affected by strong rotation gradients which can complicate their inner structures.

Asteroseismology is the only method to probe the interiors of these stars, thanks to the fact that the excited pulsational instabilities depend strongly on the stellar density profile (Suran et al. \cite{sur01}). Since the higher-order frequency spacings for pre- and post-main sequence stars are different (Suran et al. \cite{sur01}), the evolutionary phase of a field star may be determined from its pulsational eigenspectrum. The first successful seismic models of PMS stars have already been obtained (e.g., Guenther et al. \cite{gue07}; Zwintz et al. \cite{zwi07}).

In this paper we discuss MOST spacebased observations of the pulsating Herbig Ae star \hd\,\,and present a first asteroseismic model for this star.

\section{\hd}

\begin{figure*}[htb]
\centering
\includegraphics[width=\textwidth]{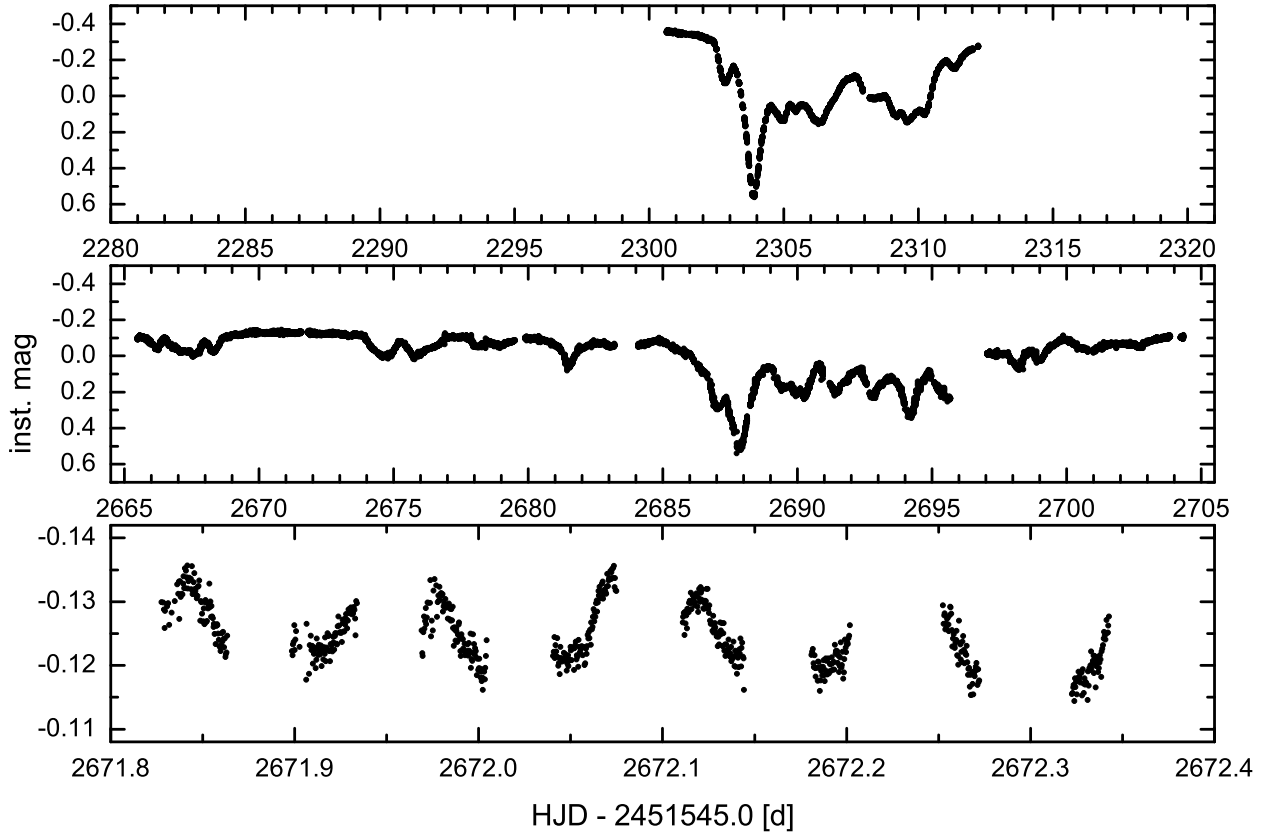}
\caption{MOST time series photometry of \hd: 2006 observations (top panel) and 2007 observations (middle panel) on the same scale. A zoom into the 2007 light curve shows the pulsational variability (bottom panel).}
\label{lcs}
\end{figure*}

The intermediate-mass Herbig Ae star \hd \,\,(V 1026 Sco) has been studied frequently in the past. Its irregular brightness variations, accompanied by color changes (i.e., the star becomes redder when fainter), can be explained by a circumstellar dust disk seen nearly edge-on (e.g., Meeus et al. \cite{mee98}) with grains smaller than $\sim$1 $\mu$m (Lecavelier des Etangs et al. \cite{les05}).  When the dust clouds are in the observer's line-of-sight, they obscure the star and cause the reddening. This phenomenon of `non-periodic Algol-like brightness minima' (also called ``UX Ori type variations'') is typical for Herbig Ae/Be stars (Grinin et al. \cite{gri94}). Even at maximum brightness \hd \,\,shows a significant amount of reddening (Meeus et al. \cite{mee98}), which indicates that the light from the star is always diluted by its dusty circumstellar environment. Note that this also makes difficult a reliable determination of the absolute magnitude and, hence, the luminosity of the star.

\hd \,\,is a Type II Herbig Ae star; i.e., it shows a double-peaked H$_{\alpha}$ emission profile (e.g. Vieira et al. \cite{vie03}). Its projected rotational velocity, \vsini, was determined to lie between 54$\,kms^{-1}$ (Meeus et al. \cite{mee98}) and 97$\,kms^{-1}$ (Vieira et al. \cite{vie03}).  With a spectral type of A8 Ve and log \Teff = 3.857 (Vieira et al. \cite{vie03}), \hd\,\, is perfectly suited to search for, and investigate, PMS pulsation.

A single pulsation frequency of 21.43 $d^{-1}$ (i.e., period of 1.12 hours) was discovered by Kurtz \& M\"uller  (\cite{kur01}) based on 6.6 hours of observations obtained during a single night. They identify this frequency as a fundamental mode according to the calculation of the pulsation constant, Q. For a detailed pulsational analysis and asteroseismic study, longer time series of data with high time sampling and excellent photometric precision were essential.  Hence, \hd\,\, was selected as a MOST Primary Science Target.

\section{MOST observations}

The MOST space telescope (Walker et al. \cite{wal03}) was launched on 30 June 2003 into a polar Sun-synchronous circular orbit of altitude 820 km. (MOST's orbital period is 101.413 minutes, corresponding to an orbital frequency of $\sim$14.2 $d^{-1}$.)  From its orbital vantage point, MOST can obtain uninterrupted observations of stars located in its Continuous Viewing Zone (CVZ) for up to 8 weeks.  The MOST satellite houses a 15-cm Rumak-Maksutov telescope feeding a CCD photometer through a single, custom broadband optical filter (covering wavelengths from 350 to 750\,nm).

MOST can supply up to three types of photometric data simultaneously for multiple targets in its field.  The mission was originally intended only Fabry Imaging, in which an in-focus image of the entrace pupil of the telescope - illuminated by a bright target star ($V < 6$) - is projected onto the instrument's Science CCD by a Fabry microlens (see Reegen et al. \cite{ree06} for details).  After MOST was operating in orbit, the pointing performance of the satellite was improved so much that a new mode of observing, Direct Imaging, was made practical.  Direct Imaging is much like conventional CCD photometry, in which photometry is obtained from defocussed images of stars in the open area of the CCD not covered by the Fabry microlens array field stop mask.  In the original mission, no scientific information was available from the guide stars used for the ACS (Attitude Control System), but now precise photometry is possible for these stars as well (see, e.g., Walker et al. \cite{wal05} and Aerts et al. \cite{aer06}).

Due to its brightness of $V = 8.81$ mag, the PMS pulsator \hd\,\, is best suited for the Direct Imaging mode of photometry.  The star lies slightly outside the MOST CVZ, so it can be observed only for a part of each 101-min orbit.  But even this time coverage represents an extremely high duty cycle compared to what is obtained with groundbased measurements. On-board exposures are 1.5 seconds long (to satisfy the cadence of guide star ACS operations), but 14 consecutive exposures are "stacked" on board to produce integrations 21 sec long, sampled about 3 times per minute.

MOST observed \hd\,\, for the first time during 22 April $-$ 4 May 2006, for 11.5 days (see Figure \ref{lcs} top panel) with a duty cycle of 34\%. (Note that the duty cycle represents the fraction of each 101-min orbit covered by the data; there are no long gaps in the time series.) The irregular UX Ori-type variations with amplitudes of nearly one magnitude due to the circumstellar dust disk are clearly visible in the light curve.

\hd\,\, was reobserved by MOST during 18 April $-$ 28 May 2007 (see Figure \ref{lcs} middle panel) with a duty cycle of 39\%. The total time span of the 2007 observations ($\sim$39 days) is more than 3 times that of the 2006 data. The gap in the light curve between $t=2696\,d$ and $t=2697\,d$ results from an interruption of the \hd\,\, observations for a high-priority MOST target of opportunity. The bottom panel in Figure \ref{lcs} is an enlargement of the 2007 light curve where the pulsational variations at the millimagnitude level can be seen clearly.

\section{Data Reduction \& Frequency Analyses}

\begin{figure*}[htb]
\centering
\includegraphics[width=0.9\textwidth]{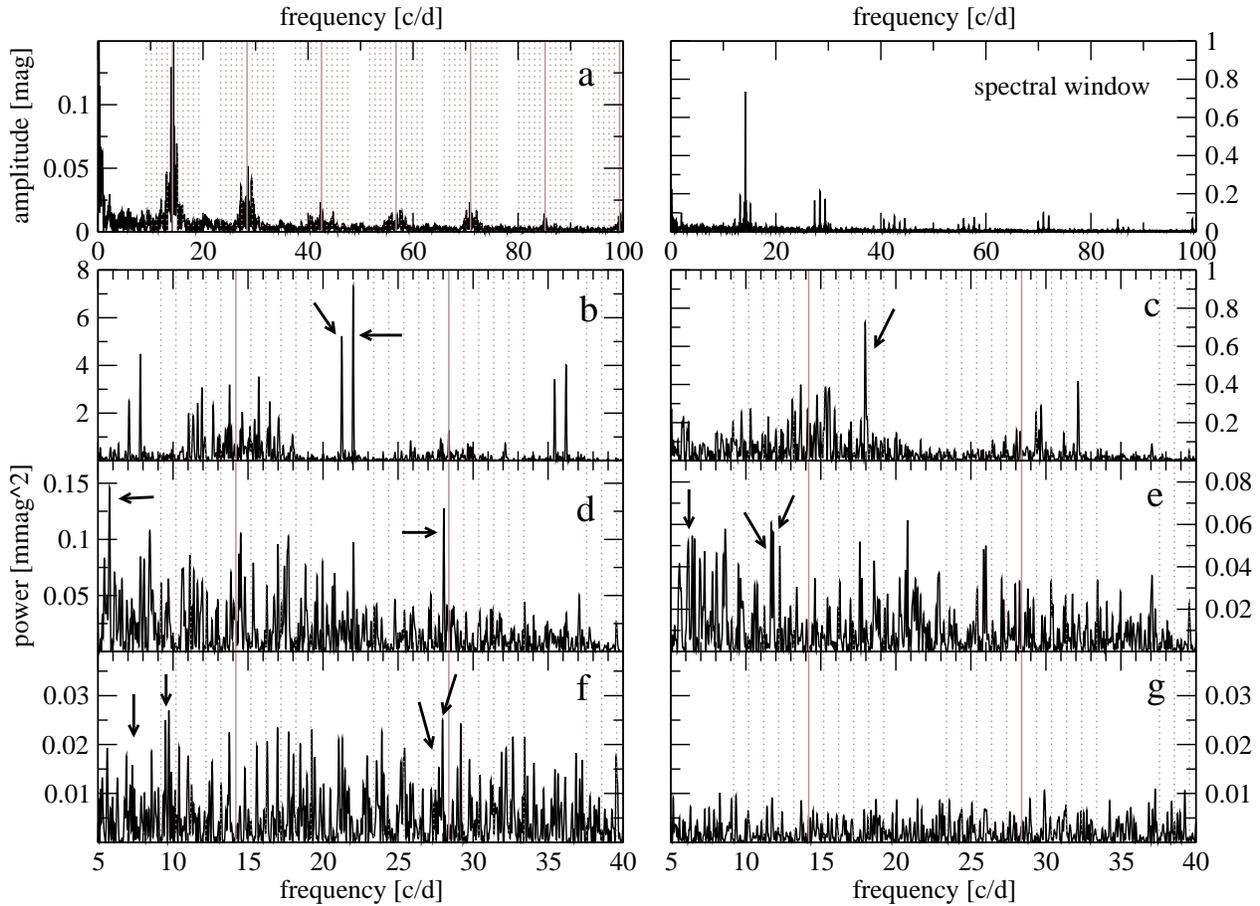}
\caption{Frequency analyses of the 2006 MOST photometry of \hd.  The amplitude spectrum is shown at top left (labelled "a") and the spectral window at top right. Panels "b" to "f" show power spectra where the identified pulsation frequencies are marked with arrows. Panel "g" (bottom right) is the residual power spectrum after prewhitening of all 151 formally significant frequencies in the data. The solid grey lines
mark the MOST satellite orbital frequency and its harmonics. The dotted grey lines are $1d^{-1}$ sidelobes of these frequencies. Note the different scalings on the y-axes.}
\label{pow2006}
\end{figure*}

For reduction of MOST Direct Imaging photometry, two different methods were developed. The first, by Rowe et al. (\cite{row06a}; \cite{row06b}), combines classical aperture photometry and point-spread function fitting to the Direct Imaging subrasters. An independent data reduction pipeline for spacebased open field photometry, which also includes automated corrections for cosmic ray hits and a stepwise pixel-to-pixel decorrelation of stray light effects on the CCD, was developed by Huber \& Reegen (\cite{hub08}). Both routines were applied to both sets of \hd\,\, photometry, and no significant differences in the qualities of the reduced light curves could be identified.  In the following analyses, the plots show data reduced using the Rowe et al. (\cite{row06a}; \cite{row06b}) method.

The main challenge in the frequency analysis of \hd\,\, was the careful distinction of the large-amplitude variations caused by the circumstellar material from the small-amplitude pulsations. {\sc SigSpec} (Reegen \cite{ree07}) and \cind \, (Reegen et al. \cite{ree08}) were used in combination to identify the pulsation frequencies.

{\sc SigSpec} (Reegen \cite{ree07}) computes significance levels for amplitude spectra of time series with arbitrary time sampling. The probability density function of a given amplitude level is solved analytically and the solution includes dependences on the frequency and phase of the signal.

\cind \,(Reegen et al. \cite{ree08}) is based on the principles of {\sc SigSpec} and compares sets of frequencies from a target and one or more comparison light curves.  A frequency is considered to be present in all data sets (and, hence, to be instrumental) if a target peak appears at the same frequency (within the frequency resolution) with at least the same significance as in the comparison data set(s). Such a coincident peak is then assigned a negative conditional significance.

For each of the two \hd\,\, light curves and their corresponding background light curves, amplitude and significance spectra were computed. The results of these calculations were then submitted to \cind\,\, separately for each year of the observations, to identify coinciding peaks between the target and the background. Such peaks can originate from the background itself (due to scattered Earthshine variations) or from instrumental effects. The irregular light variations of \hd\,\, introduce a `pseudo-periodicity' and an excess of significant peaks in the low-frequency domain (i.e., from 0 to 3 $d^{-1}$) which can be seen, e.g., in Figure \ref{results}. The frequency range where $\delta$-Scuti-like pulsations are expected $-$ i.e., from $\sim$5 to 70 $d^{-1}$ $-$ is well separated from the peaks caused by the irregular variability.  Nevertheless, the formally significant frequencies in this domain have to be interpreted carefully for several reasons.

First, alias frequencies from the low-frequency domain can appear at higher frequencies due to the spectral window. A manual check against such spurious peaks as well as a dedicated test (see Section \ref{test}) were carried out. Second, instrumental frequencies related to the orbit of the satellite, its harmonics and 1$d^{-1}$ sidelobes are present in the range of the expected pulsation frequencies. All significant peaks that can be related to one of the above mentioned instrumental frequencies within the frequency resolution (computed according to Kallinger et al. \cite{kal08}) were discarded. Finally, the identified pulsation frequencies need to be significant in the analyses of both \hd\,\, time series.

\subsection{2006 data}

The top left panel (labelled "a") in Figure \ref{pow2006} shows the amplitude spectrum of the 2006 data from 0 to 100 $d^{-1}$ where the solid grey lines mark the orbital frequency of MOST and its multiples and the dotted grey lines represent the corresponding 1$d^{-1}$ sidelobes. The spectral window is shown in the top right panel.

We find 151 peaks (grey lines in the top layer of Figure \ref{results}) to be formally significant; i.e., their significance criterion is larger than 5 (Reegen \cite{ree07}) which corresponds to an amplitude signal-to-noise (S/N) \mbox{ratio $>$ 4} (Breger et al. \cite{bre93}; Kuschnig et al. \cite{kus97}). About half of these frequencies lie between 0 and 3 $d^{-1}$, and can be attributed to the irregular variability caused by the dusty circumstellar environment. Another large fraction of the formally significant peaks can be related to the orbital frequency of the satellite, its harmonics and $1d^{-1}$ sidelobes.

In panel "b" of Figure \ref{pow2006}, the appearance of alias frequencies due to the gaps in the data per MOST orbit can be seen: aliases of the two highest pulsation frequencies at 21.250 $d^{-1}$ and 22.009 $d^{-1}$ appear at $\pm 14.2 d^{-1}$ which is the orbital frequency of MOST. After prewhitening these two main frequencies, the respective aliases also disappear.

The residual noise level in the 2006 data set is 38 ppm.

\subsection{2007 data}

\begin{figure*}[htb]
\centering
\includegraphics[width=0.9\textwidth]{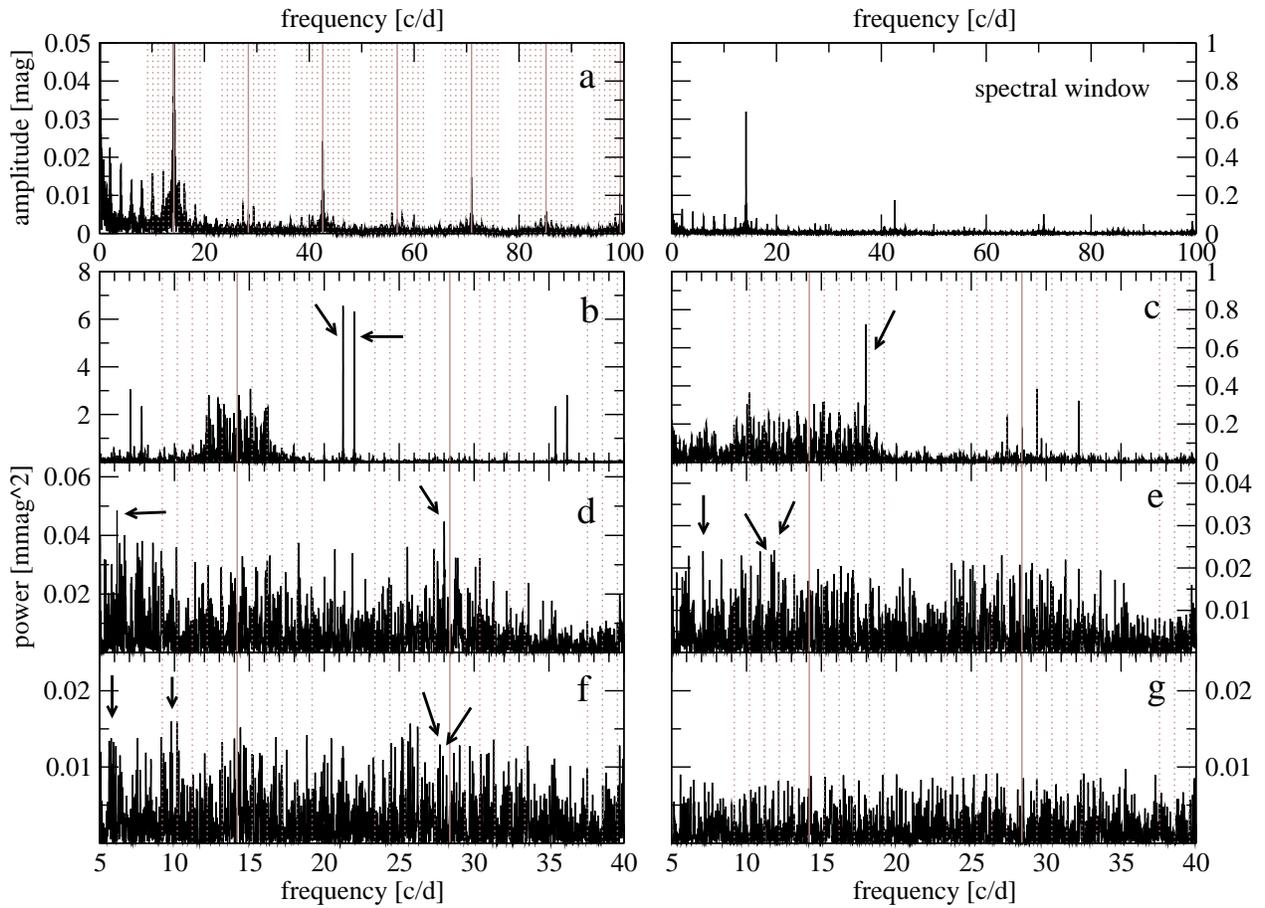}
\caption{Frequency analysis of the 2007 MOST observations of \hd. The format of the figure is the same as Figure 1. Panel "g" (bottom right) is the residual spectrum after prewhitening of all 415 formally significant frequencies found in this time series.  Note the different scalings on the y-axes.}
\label{pow2007}
\end{figure*}

The top left panel (labelled "a") in Figure \ref{pow2007} shows the amplitude spectrum of the 2007 data from 0 to 100 $d^{-1}$.  Since the 2007 MOST time series is more than 3 times longer than the 2006 data set, the amplitude spectrum of the 2007 photometry has an obviously lower noise level.

\begin{figure}[htb]
\centering
\includegraphics[width=7.5cm]{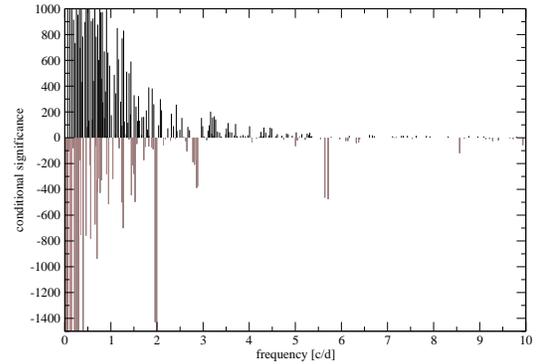}
\caption{A closer look at the low-frequency domain of the results of the 2007 data analysis with \cind. The $2-d^{-1}$ frequency is known to be due to the gaps introduced in the \hd\,\, time series. This and other peaks with negative values of their conditional significance are shown by grey lines. Peaks that appear only in the stellar light curve have positive conditional significance (black lines).}
\label{cind07}
\end{figure}

The regularly spaced peaks between 0 and 14 $d^{-1}$ visible in the amplitude spectrum are also present in the spectral window (top right in Figure \ref{pow2007}). The cause is gaps in the data due to interruptions to observe yet another high-priority MOST target of opportunity, monitored every half-day during the 2007 \hd\,\, run. This regular spacing of 2 $d^{-1}$ and its harmonics appear in the amplitude spectrum of the background light curve with even higher amplitudes. When submitting the star and the background light curves to \cind, these frequencies get negative conditional significances and are therefore rejected. Figure \ref{cind07} shows the results of the analysis with \cind.

\begin{figure}[htb]
\centering
\includegraphics[width=7.5cm]{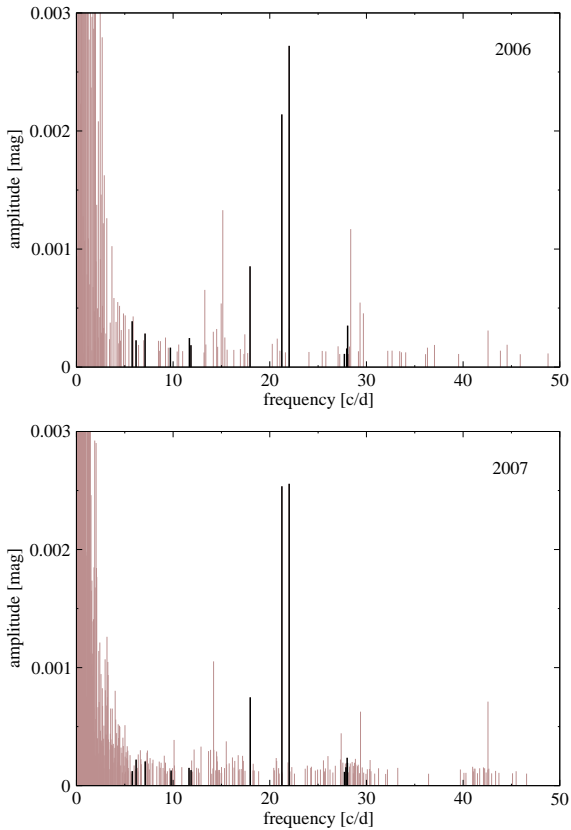}
\caption{Significance spectra of the 2006 data (top) and 2007 data (bottom). Grey lines mark all formally significant peaks; black lines are the 12 identified pulsational frequencies common to both data sets.}
\label{results}
\end{figure}

There are 451 formally significant frequencies in the 2007 MOST data (grey lines in the bottom panel of Figure \ref{results}), where a third of the peaks lie between 0 and 3 $d^{-1}$ and are due to the irregular light variations of the star. As with the 2006 data, the alias frequencies due to the spectral window can be seen at $\pm 14.2 d^{-1}$ around 21.252 $d^{-1}$ and 22.008 $d^{-1}$ (see Figure \ref{pow2007}, panel "b").

The residual noise level in the 2007 data set is 30 ppm.

\subsection{Pulsational frequencies}

From the numerous formally significant peaks in the amplitude spectra of \hd, 12 were identified to likely originate from pulsation because they meet the following criteria: (a) They are not related to the
orbital frequency of the satellite, its harmonics or 1$d^{-1}$ sidelobes. (b) They appear significantly in the data sets from both years. (c) They cannot be attributed to the peaks introduced by the irregular light variations and their alias frequencies. (d) They passed the test described below in \ref{test}.

The 12 frequencies, their amplitudes and significances are listed in Table \ref{freqs}, where the errors given in parentheses are computed using the formulae given by Kallinger et al. (\cite{kal08}). Although the significances and corresponding amplitudes of some of the listed frequencies are quite low, they all meet the criteria described above.

\begin{table*}[htb]
\caption{Pulsation frequencies of \hd \,\,identified in both data sets (2006 and 2007), their amplitudes and significances, where the respective frequency errors are given in parentheses. The table is sorted by increasing frequency, where the frequencies are numbered according to their significances.}
\label{freqs}
\begin{center}
\begin{footnotesize}
\begin{tabular}{rrrrrcrrrrc}
\hline
\multicolumn{1}{c}{\#} & \multicolumn{4}{c}{2006 data}  & \multicolumn{4}{c}{2007 data}  \\
\hline
 & \multicolumn{2}{c}{freq} & \multicolumn{1}{c}{amp} & \multicolumn{1}{c}{sig} &
\multicolumn{1}{c}{Fig. \ref{pow2006}} & \multicolumn{2}{c}{freq} & \multicolumn{1}{c}{amp} & \multicolumn{1}{c}{sig} & \multicolumn{1}{c}{Fig. \ref{pow2007}} \\
 & \multicolumn{1}{c}{[$d^{-1}$]} & \multicolumn{1}{c}{[$\mu$Hz]} & \multicolumn{1}{c}{[mmag]} & & \multicolumn{1}{c}{layer}  & \multicolumn{1}{c}{[$d^{-1}$]} & \multicolumn{1}{c}{[$\mu$Hz]} & \multicolumn{1}{c}{[mmag]} & & \multicolumn{1}{c}{layer} \\
\hline
f11  &  5.77(1) & 66.7(2) & 0.388 & 39.4 & d & 5.78(1) & 66.8(1) & 0.123 & 6.8 & f \\
f4  &  6.17(2) & 71.4(3) & 0.227 & 16.3 & e & 6.17(1) & 71.4(1) & 0.220 & 15.1 & d \\
f6  &  7.10(3) & 82.2(3) & 0.284 &  9.5 & f & 7.11(1) & 82.3(1) & 0.205 & 11.3 & e \\
f10 &  9.72(3) & 112.5(3) & 0.164 &  9.8 & f & 9.79(1) & 113.3(1) & 0.128  &  7.7 & f \\
f7 & 11.68(2) & 135.2(2) & 0.246 & 21.2 & e & 11.65(1) & 134.8(1) & 0.150 & 10.2 & e  \\
f8 & 11.85(3) & 137.1(2) & 0.185 & 11.8 & e & 11.86(1) & 137.2(1) & 0.131 & 11.1 & e   \\
f3 & 17.96(1) & 207.8(1) & 0.854 & 105.5 & c & 17.979(2) & 208.02(2) & 0.748 & 136.6 & c  \\
f2 & 21.25(1) & 245.86(7) & 2.141 & 228.9 & b & 21.252(1) & 245.89(1) & 2.535 & 384.3 & b  \\
f1 & 22.01(1) & 254.64(6) & 2.721 & 263.8 & b & 22.008(1) & 254.64(1) & 2.557 & 400.2 & b  \\
f12 & 27.72(4) & 320.8(4) & 0.112 & 5.8 & f & 27.72(1)  & 320.8(1) & 0.116 & 6.2 & f  \\
f9  & 27.96(3) & 323.5(3) & 0.158 & 10.2 & f & 27.88(1)& 322.6(1) & 0.156 & 10.7 & f \\
f5 & 28.05(1) & 324.6(2) & 0.351 & 35.4 & d & 28.04(1) & 324.4(1) & 0.193 & 14.7 & d \\
\hline
\end{tabular}
\end{footnotesize}
\end{center}
\end{table*}

\subsection{Test with 2007 subset light curve}
\label{test}

\begin{figure}[htb]
\centering
\includegraphics[width=7.5cm]{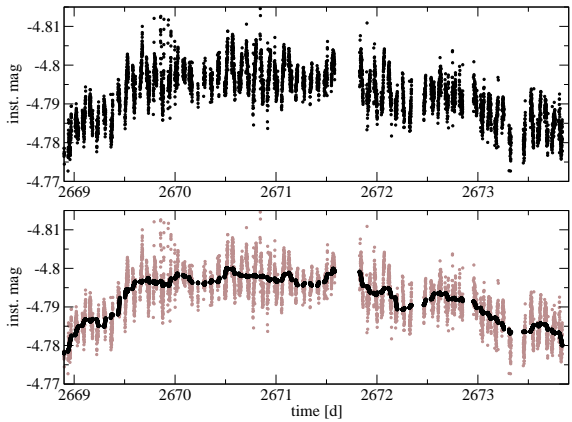}
\caption{Five-day subset light curve from the 2007 data set (top) and a moving average fit over a time interval of 4 hours (in black) plotted over the light curve (bottom).}
\label{cutlc}
\end{figure}

The reliability of the selected 12 pulsation frequencies was investigated in the following way. The longest part of the light curves with minimal irregular light variations lies in the 2007 data set from $t=2668.9\,d$ to $t=2673.9\,d$ (see top panel in Figure \ref{cutlc}).

First, a frequency analysis of this 5-day subset light curve was carried out with \SigSpec. Although this is the "quietest" part of the light curve, irregular variability is still clearly present. To reproduce the shape of the irregular light variations in this data subset, moving averages were computed.  Using a time interval of 4 hours for the boxcar allows one to smear out the pulsation effects deliberately, without significant suppression of the irregular variations. The bottom panel of Figure \ref{cutlc} shows the subset light curve (grey dots) where the boxcar is overplotted (black dots).

A frequency analysis of the moving average was then computed with {\sc SigSpec}. The results of the subset light curve and the boxcar light curve were then compared using \cind. If frequencies are found in both data sets (i.e., they have negative conditional significances in the output of \cind), they very likely originate from the circumstellar environment and are not caused by pulsation.

\begin{figure}[htb]
\centering
\includegraphics[width=7.5cm]{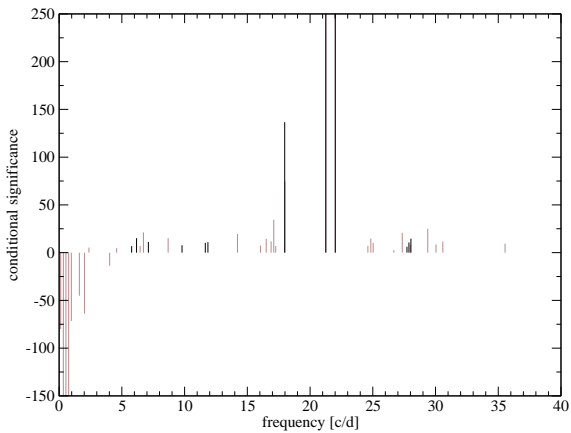}
\caption{Comparison between the frequencies of the 5-day subset light curve and the moving average fit over a time interval of 4 hours using \cind. The grey lines show the output of \cind, where peaks with negative conditional significances appear in both data sets, and are unlikely to be caused by pulsation. The 12 pulsation frequencies identified in \hd \,\,have positive conditional significances. Their positions are shown as black lines for better visibility.}
\label{cutlc-cind}
\end{figure}

All 12 frequencies previously attributed to pulsation were {\it only} present in the computation of the subset light curve. They do {\it not} show up in the frequency analysis of the boxcar and have positive conditional significances in \cind\,\, (see Figure \ref{cutlc-cind}).

\section{Asteroseismic Analysis}

\subsection{Model grid}

We compare the observed frequencies from the 2007 run for \hd\,\, to asteroseismic models in order to determine the nature of the pulsations and to constrain the star's effective temperature, luminosity and mass independently. We note that the model fits to the 2006 frequencies (and averaged 2006 and 2007 frequencies) are nearly identical, which is as expected since the differences between corresponding frequencies is less than 1.0 $\mu$Hz, and typically less than 0.25 $\mu$Hz.

Grids of models were constructed using the Yale Rotating stellar Evolution Code (YREC; Guenther et al. \cite{gue92}). PMS models were evolved from the Hayashi track (Hayashi \cite{hay61}) before deuterium burning to the ZAMS. The grids include models with masses ranging from 1.00 to 5.00\,\Msun in steps of 0.01$\,M_{\odot}$. Each evolutionary track is resolved into approximately 1000 models. A near-solar composition (Z = 0.02, Y = 0.27) was assumed.

The constitutive physics of the models are current and include OPAL98 (Iglesias \& Rogers \cite{igl96}), the Alexander \& Ferguson (\cite{ale94}) opacity tables, and the Lawrence Livermore National Laboratory equation of state tables (Rogers \cite{rog86}; Rogers et al. \cite{rog96}). The mixing length parameter used to describe the temperature gradient in convective regions according to the B\"ohm-Vitense (\cite{boe58}) mixing-length theory, was adjusted from calibrated solar models. The PMS evolutionary tracks used in this analysis compare well with the tracks of D'Antona \& Mazzitelli (\cite{dan94}). Our models do not include mass accretion (Palla \& Stahler \cite{pal92}, \cite{pal93}), which may occur prior to initial deuterium burning. We used Guenther's nonradial stellar pulsation code (Guenther \cite{gue94}) to compute the adiabatic pulsation spectra.

We used a dense and extensive grid of model spectra to find the best match to the observed frequency spectrum. This method was originally developed by Guenther \& Brown (\cite{gue04a}) for modeling stars in more advanced evolutionary stages where mode bumping complicates the search for the best match to the observed oscillations. The quality of the match is quantified by the simple $\chi^2$ relation as described, e.g., in Guenther et al. (\cite{gue07}), that compares the model and observed frequencies weighted by the model and observational uncertainties. The grid of oscillation spectra is searched to locate local minima in chi-squared. A value of $\chi^2 \le 1$ means that the difference between the observed frequencies and the corresponding nearest model frequency is less than the rms of the model and observational uncertainties. We assumed the observed frequencies have an uncertainty of $\pm 0.3 \mu$Hz (corresponding to the reciprocal of the 2007 observing run duration).

\subsection{Modelling \hd}

Before trying to find a model whose oscillation spectrum matches an observed set of frequencies, we first preview the observed frequencies in an echelle diagram. This enables us to quickly check if the observed frequencies show the ridge-like structure expected for p-modes. The frequencies of p-modes, especially those of PMS stars whose chemical compositions are nearly homogeneous, are relatively uniformly spaced. Specifically, we expect modes of similar $l$-value to be separated by the large spacing, defined as the frequency difference between $n$ and $n+1$ p-modes (of similar $l$-value.) By plotting the frequency of a mode versus its frequency modulo the large spacing in an echelle diagram, one can immediately identify sequences of frequencies or ridges that correspond to similar $l$-valued modes. The large spacing is  approximately equal to the characteristic frequency spacing (Tassoul \cite{tas80}) which can be computed directly from a stellar model of the star.

\begin{figure}[htb]
\centering
\includegraphics[width=0.5\textwidth]{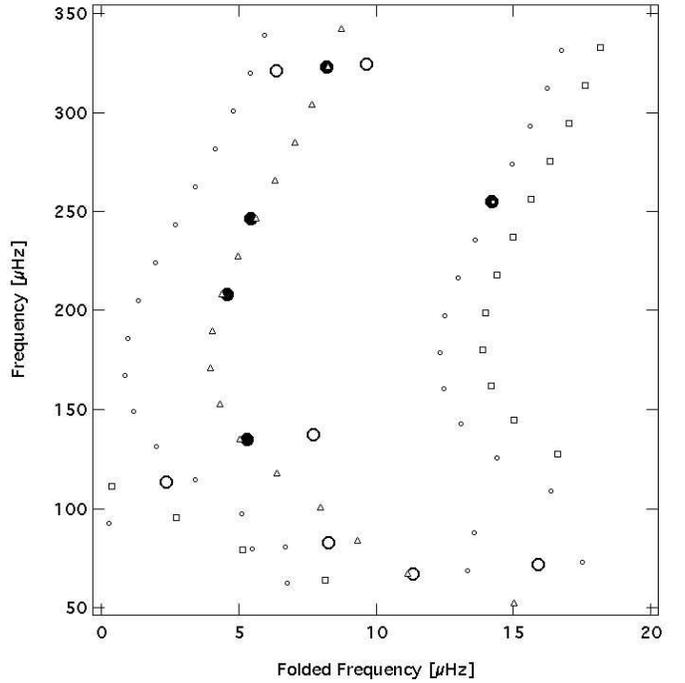}
\caption{Echelle diagram of the model from the best model fit to the five frequencies f1, f2, f3, f7, and f9. Fitted frequencies are plotted as large filled circles, frequencies not used in the fit as large open circles. $l$ = 0 modes are shown with small open squares, $l$ = 1 modes as small open triangles, $l$ = 2 and 3 modes with small open circles running parallel to the $l$ = 0 and 1 modes, respectively.}
\label{echelle}
\end{figure}

We found that no single folding frequency for the echelle diagram allows all 12 observed frequencies to fall onto $l$-ridges. Indeed, it was immediately clear, based on their proximity, that the two frequencies f7, f8 and the three frequencies f5, f9, and f12 appear to be rotationally split modes.  HD 142666 has a projected rotational velocity \vsini of about 50-100\,$kms^{-1}$, which is consistent with this interpretation. Because our model frequencies are obtained from a non-rotating stellar model, no rotationally split mode frequencies are included in our grid search. Hence, the oscillation spectrum searching algorithm will fail to fit, or be skewed by, any observed frequency that is an $m \ne 0$ (i.e., not the central) split mode. Consequently, we included only the central frequency, f7, and f9. The rotationally split assumption is, of course, speculative. Regardless of their true nature, as can be seen in the echelle diagram (Figure \ref{echelle}), we are unable to find any model fits to all these frequencies simultaneously since their spacings does not match simultaneously the small and large spacings, as would be required to fit all three frequencies.

\begin{figure}[htb]
\centering
\includegraphics[width=0.5\textwidth]{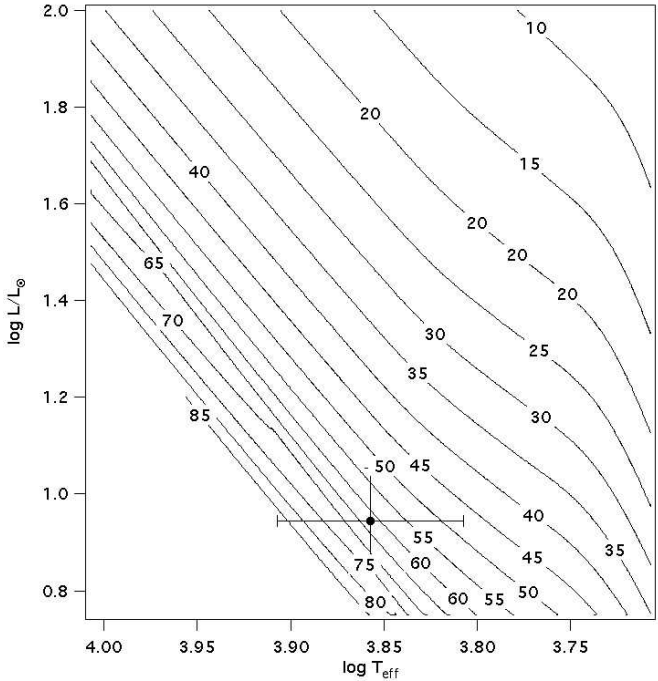}
\caption{Contour plot of the characteristic frequency spacing (in $\mu$Hz) near the star's position in the HR-diagram showing that the large spacing is expected to be between 45 and 80 $\mu$Hz.}
\label{delnu}
\end{figure}

Although the ultimate goal of the seismic analysis is to determine a model of \hd\,\, based on observed frequencies alone, it is extremely useful to have some prior information about the star. Unfortunately, because \hd\,\, is surrounded by a circumstellar disk, its location in the HR-diagram is uncertain. The effective temperature of \hd\,\,as found in the literature varies wildly.  For example, Vieira et al. (\cite{vie03}) find \Teff = 7200\,K while Dominik et al. (\cite{dom03}) find \Teff = 8500\,K for this star. The recent analysis of data obtained with the ESPADONS spectrograph (J. Grunhunt \& G. Wade, priv. comm.) yielded \Teff  = 7500\,K. Here we adopt for reference the value of Vieria et al. (\cite{dom03}), log \Teff = 3.857$\pm$0.05. The only estimate of luminosity we could find in the literature is by Monnier et al. (\cite{mon05}) who obtain $L = 8.8\pm2.5$\,\Lsun\,\, derived from a relation between circumstellar disk radius and stellar luminosity. The result is highly uncertain, since the input parameters of the disk are, themselves, uncertain.

We computed the characteristic frequency for the models in our PMS grid that are in the vicinity of \hd's location in the HR diagram and plotted them in the contour plot shown in Figure \ref{delnu}. Figure \ref{delnu} shows that if the star is correctly positioned in the HR diagram, then we can expect the large spacing to be between 45 and 80 $\mu$Hz. As we show below, our best model fits have a much smaller large spacing ($\sim$15 $\mu$Hz).

\begin{figure}[htb]
\centering
\includegraphics[width=0.5\textwidth]{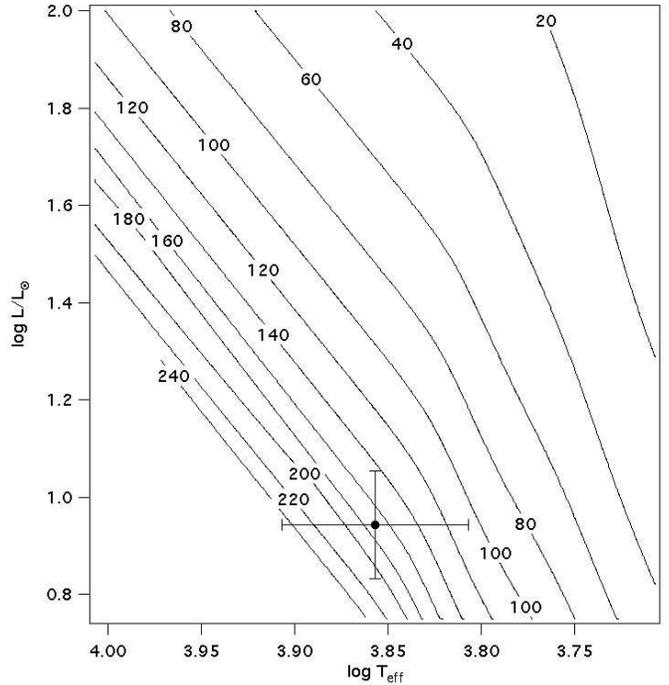}
\caption{Contour plot of the frequency (in $\mu$Hz) of the fundamental radial mode near the star's position in the HR-diagram.}
\label{fundamental}
\end{figure}

Figure \ref{fundamental} is a contour plot on which we show the frequency of the fundamental radial model for all the models in our PMS grid near \hd's presumed location in the HR diagram. If the HR diagram location of \hd\,\,is correct, then the p-modes must have frequencies greater than $\sim$100 $\mu$Hz. Consequently, without further analysis we can immediately see that, if the observed frequencies down to 67 $\mu$Hz are all intrinsic p-modes, then \hd's presumed luminosity and/or temperature are incorrect. Of course, it is also possible that, if intrinsic, the lowest frequencies are g-modes. Without corroborating oscillation observations, and with a highly uncertain HR diagram location, we cannot say for certain.

\begin{figure}[htb]
\centering
\includegraphics[width=0.5\textwidth]{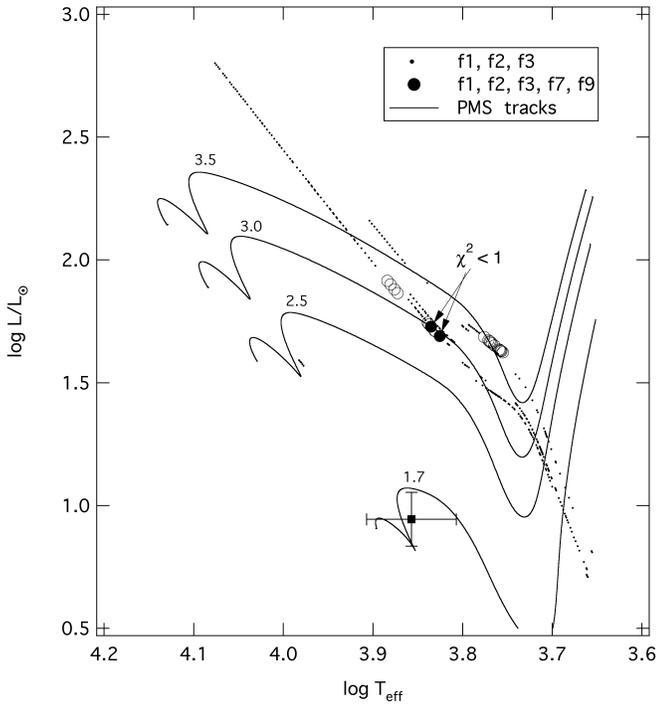}
\caption{The HR-diagram position of HD 142666 (filled square with error bars) together with all the models that have oscillation frequencies matching the combinations of f1, f2, and f3 (small filled circles) and f1, f2, f3, f7, and f9 (large open circles for $\chi^2 \le 2$ and large filled circles for $\chi^2 \le 1$) with the frequency uncertainty. PMS evolutionary tracks are also plotted to show the relative evolutionary phase of the model fits. The masses, in solar mass units, label the individual tracks. }
\label{hrd}
\end{figure}

We began our spectrum fitting analysis by considering only the three most significant frequencies f1, f2, and f3. In the HR-diagram shown in Figure \ref{hrd}, we plot all the models that have oscillation frequencies that match the three observed frequencies within the assumed frequency uncertainty of $\pm$0.3 $\mu$Hz. The model oscillation spectra include $l$ = 0, 1, and 2 p-modes. We find a variety of models spread throughout the HR-diagram that fit the three most significant frequencies with $\chi^2 \le 1$. None, though, lie near the HR diagram position of the star.

No model spectra in our PMS grid were found to fit, with $\chi^2 \le 1$, all the observed frequencies excluding f5, f8, and f12 (the frequencies tentatively identified as side lobes to rotationally split modes). We, therefore, considered the possibility that some of the frequencies are not p-modes. After investigating several combinations of frequencies, we applied the following restrictions to obtain the final five frequencies that we are able to fit. First, we considered only frequencies $>$ 125 $\mu$Hz, thus eliminating frequencies f4, f6, f10 and f11 from our fitting tests. Second, we took the central frequency only in the triplet f5, f9, f12, i.e., f9. And, finally, we chose the most significant frequency, f7, from the doublet f7, f8. Figure \ref{hrd} shows the best model fits to the resultant set of frequencies f1, f2, f3, f7, and f9. The large filled circles correspond to model fits with $\chi^2 < 1$ and the large open circles to models with $\chi^2 < 2$. The model fits to all five frequencies is a subset of the model fits to the three most significant frequencies.

Figure \ref{echelle} is an echelle diagram of the model frequencies from the best model fit, i.e., the model with the lowest $\chi^2$ ($\sim$0.3), to the selected five frequencies. The model has log \Teff = 3.8267, log $L$/\Lsun  = 1.6903, radius = 5.2\Rsun and mass = 2.96\Msun . The five fitted frequencies are indicated by large filled circles. The remaining observed frequencies are noted by large open circles. The $l$ = 0 model frequencies are shown with small open squares, the $l$ = 1 mode frequencies are shown with small open triangles, and the $l$ = 2 and 3 model frequencies are shown with small open circles running parallel to the $l$ = 0 and 1 modes, respectively. The echelle diagram shows that the five observed frequencies are well fit by the $l$ = 1 and 2 p-modes. We note that we do not at any time during the fitting process force specific frequencies to be fitted, for example, by only $l$ = 0 modes. If we consider only models that lie within the estimated effective temperature range (ignoring the luminosity discrepancy) that fit the three most significant frequencies then we arrive at nearly the same models, and produce nearly the same echelle diagrams as our best model fit to all five frequencies. Essentially, the additional two frequencies help constrain the star's location in the HR diagram to coincide with the estimated effective temperature range.

The identification of rotational splittings for the triplet f5, f9, and f12 and the doublet f7 and f8 is somewhat uncertain in that the splitting spacings are not the same for the lower frequency pair compared to the higher frequency triplet. The larger splitting for the lower frequency pair (assuming that the missing frequency is not the central frequency) implies that the core of \hd\,\,is rotating more rapidly than the surface layers. For our best fitting model a splitting on the order of 2 $\mu$Hz, under the assumption of solid body rotation, yields a surface rotation period of $\sim$6\,d and a surface rotation velocity at the equator of \mbox{$\sim$50 $kms^{-1}$.} This is a typical rate for PMS stars and compares well to the observed \vsini measurements within the uncertainties of our own velocity determination which are $\pm$50\%, owing to the uncertainty in frequency splitting itself which is on the order of 1 $\mu$Hz, i.e., 50\%.

If the five frequencies are p-modes and if our classically determined PMS models are reasonable representations of the interior structure then the observed frequencies are able to constrain the mass, luminosity, and surface temperature of \hd. The asteroseismic mass, for example, is far away from the mass that would be determined from simply fitting the star's location in the HR-diagram, i.e., $\sim$1.7\Msun (see Figure \ref{hrd}).
There are, though, several issues that need to be resolved. We need to understand why the best fitting models lie well away from the star's presumed location in the HR diagram. Possibly this could be due to the presence of the circumstellar disk, which makes photometric determinations of the star's luminosity and effective temperature highly uncertain. We also have to wonder why none of the frequencies fit are radial modes (i.e., $l$ = 0). Most likely, our mode identifications and the specific model fit are not completely correct. For example, if the outer envelope and atmosphere of the models are slightly off, a reasonable possibility considering the amount of dust in the neighborhood of the star, then the higher-frequency modes that are sensitive to the outer layers of the model, will also be off. A model uncertainty of $\pm 1 \mu$Hz is possible, which is enough to allow $l$ = 0 modes to be fit, but not enough to push the HR-diagram position of the best fitted models toward the star's apparent position. We believe a reasonable next step is to include mass accretion, as is done in the best models of Palla \& Stahler (\cite{pal92}, \cite{pal93}), and see how large is the effect and if there is any improvement in the fits.

At this time we note that adjustments to mixing length parameter, the helium abundance, and heavy element abundance will perturb the location of the models in the HR diagram only by a small amount (e.g., $\sim$300\,K) $-$ far from enough to account for the gap between the star's putative location in the HR diagram and the seismically fitted models.

\section{Conclusions}

MOST high-precision time series photometry of the Herbig Ae star \hd\,\,obtained in 2006 and 2007 was used to investigate the pulsational properties of the star. This task was complicated by the fact that \hd\,\,is surrounded by a dense circumstellar dust disk which causes irregular light variations introducing high-amplitude signal at low frequencies (i.e., from 0 to 3 $d^{-1}$). We were able to disentangle the pulsational variability with amplitudes at the millimagnitude level from the irregular variations with amplitudes of up to 200 mmag in the Fourier domain.

Although numerous peaks appear to be formally significant, we identify only 12 frequencies as pulsational in origin in the data sets of both years. Therefore, only these were used for the asteroseismic analysis.

For our model fitting, we further restricted the set of frequencies by excluding frequencies below 125 $\mu$Hz. Also, based on their proximity in frequency space, we included only the central frequency from the {f5, f9, f12} triplet, and the most significant frequency from the {f7, f8} pair, since we suspect they are rotationally split modes.  We, thus, restricted the fits to the five modes {f1, f2, f3, f7, f9}.

We are able to obtain good fits ($\chi^2 \le 1$) to the frequencies for $l$ = 0, 1 and 2 p-modes. Significantly, though, all the model fits lie well outside the HR diagram uncertainty box for \hd. Even when we fit only the three most significant frequencies (f1, f2, and f3), the models with $\chi^2 \le 1$ lie far away from the presumed location of the star in the HR diagram. Furthermore, we note that the fundamental p-mode frequencies in the vicinity of \hd's location in the HR diagram are higher than the lowest frequencies observed.  Higher luminosity models, though, do have lower fundamental frequencies.

The failure to find a model that fits even the three most significant modes that at the same time lies within the uncertainty box for the star's location in the HR diagram leads us to suggest three areas to investigate further:

\noindent
1. The models are too crude and need additional modeling physics such as mass accretion during the pre-deuterium burning phases to fit the frequencies and the star's location in the HR diagram simultaneously. The presence of the circumstellar disk supports this idea but the size of the effect on p-mode frequencies needs to be determined.

\noindent
2. The third mode in the possibly rotationally split group {f7, f8} needs to be identified to help support the hypothesis that the modes are rotationally split. This would also help mode identification in that the modes would then have to be nonradial.

\noindent
3. The luminosity of the star needs to be better constrained. This would rule out either the model fits or the p-mode nature of the lowest frequencies observed. If the star's luminosity is correct then the lowest observed frequencies are either non-intrinsic or they are g-modes. Note that for \hd \,\,no HIPPARCOS (ESA \cite{hip97}) parallax is available and the error of the parallax from the ASCC 2.5 Catalog (Kharchenko \cite{kha01} Kharchenko et al. \cite{kha05}) is a factor of three larger than the parallax itself.

\begin{acknowledgements}
K.Z., T.K., M.G., R.K. and W.W.W. acknowledge support by the Austrian {\it Fonds zur F\"orderung der wissenschaftlichen Forschung} (KZ: project T335-N16; TK, MG, RK and WWW: project P17580).
The Natural Sciences and Engineering Research Council of Canada supports the research of D.B.G., J.M.M., A.F.J.M., S.M.R. and M.P.C; A.F.J.M. is also supported by FQRT (Qu\'ebec), and R.K. is also supported by the Canadian Space Agency.
Special thanks goes to Gregg Wade and Jason Grunhunt who provided the newest values for \Teff.
\end{acknowledgements}

\end{document}